\documentclass[aps,twocolumn,amsmath,amssymb,showpacs,prb,floatfix]{revtex4-1}
\usepackage{amsmath}
\usepackage{graphicx}
\usepackage{epsfig}
\newlength{\upit}\upit=0.1truein

\newcommand{\ltappr}{{{\lower4pthbox{$<$} } \atop \widetilde{ \ \ \ }}}
\newlength{\bxwidth}\bxwidth=1.5 truein

\newcommand{\dg}{^{\dagger }}

\newcommand \bea {\begin{eqnarray} }
\newcommand \eea {\end{eqnarray}}
\newcommand{\bk}{{\bf{k}}}
\newcommand{\bq}{{\bf{q}}}

\newcommand{\ba}{{\bf{a}}}

\newcommand{\mot}{LiZn$_2$Mo$_3$O$_8$\;}
\newcommand{\motp}{LiZn$_2$Mo$_3$O$_8$}

\newlength{\figwidth}
\newlength{\shift}
\newcommand{\fg}[3]
{
\begin{figure}[ht]

\vspace*{-0cm}
\[
\includegraphics[width=\figwidth]{#1}
\]
\vskip -0.2cm
\caption{\label{#2}
\small#3
}
\end{figure}}

\begin{document}

\title{Emergent honeycomb lattice in \mot} 

\author{Rebecca Flint and Patrick A. Lee} 
\affiliation{ 
Department of Physics, Massachusetts Institute of Technology, Cambridge, Massachusetts 02139,
U.S.A.  }
 
\begin{abstract} 
We introduce the idea of \emph{emergent lattices}, where a simple lattice decouples into two weakly-coupled lattices as a way to stabilize spin liquids.  In \motp, the disappearance of 2/3rds of the spins at low temperatures suggests that its triangular lattice decouples into an emergent honeycomb lattice weakly coupled to the remaining spins, and we suggest several ways to test this proposal.  We show that these orphan spins act to stabilize the spin-liquid in the $J_1-J_2$ honeycomb model and also discuss a possible 3D analogue, Ba$_2$MoYO$_6$ that may form a ``depleted fcc lattice.''
\end{abstract}
 
\maketitle

Spin liquids are highly correlated magnetic states that break no symmetries and
hold the theoretical promise of new fractionalized excitations and topological orders\cite{lee08, balents10}.  Realizing spin liquids experimentally is a hard problem, although we have a few recent examples on the triangular\cite{tri,tri2} and kagom\'e\cite{kagome} lattices.  To explore the full range of possible spin-liquids, we would like to realize spin liquids on a wide variety of lattices and having an additional method to stabilize the spin liquid phase would be extremely helpful.  In this paper, we show how forming a low temperature \emph{emergent honeycomb lattice} out of the triangular lattice can stabilize the spin liquid state, and discuss the relevance of this idea to \motp.  

Despite its bipartite nature, the low coordination number ($z = 3$) of the honeycomb lattice increases the quantum fluctuations, and numerical studies have suggested that a spin-liquid region can emerge out of the N\'eel state with decreasing $U$ (Hubbard model)\cite{meng10} or increasing next-nearest neighbor coupling, $J_2$ (Heisenberg model)\cite{lu11,wang10,clark11,gong13}. Although further studies now suggest weak magnetic order in the Hubbard model\cite{sorella12,clark13} and the existence/size of the spin liquid region in the Heisenberg model are controversial\cite{albuquerque11,ganesh13,zhu13}, the energy of the spin liquid is clearly competitive. Currently there are no experimental examples of honeycomb spin liquids, but the triangular lattice material, \mot [\onlinecite{sheckelton12}] might provide an unexpected realization, as it could deform into an emergent honeycomb lattice weakly coupled to orphaned central spins.  

\fg{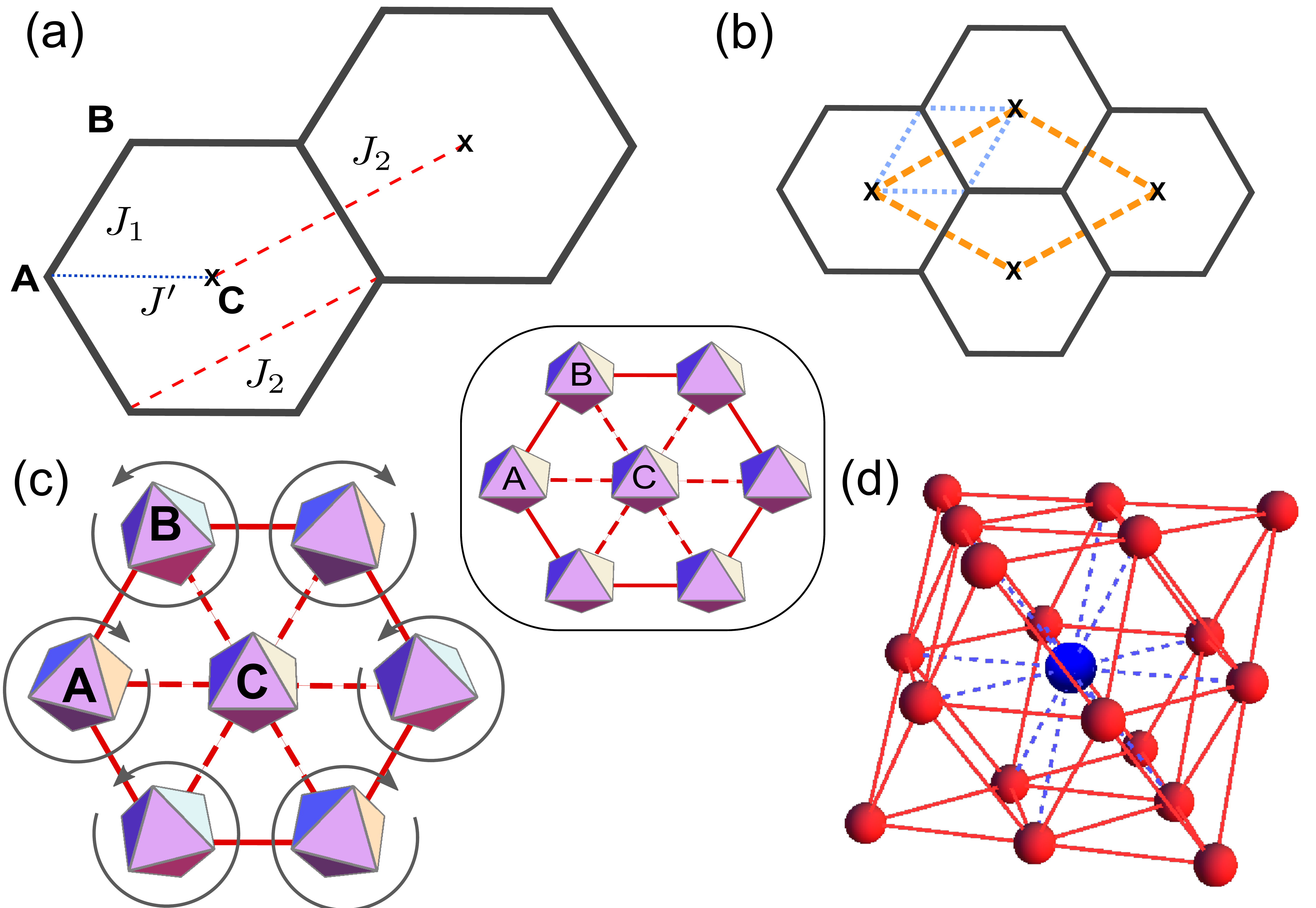}{lattice}{(a) $J_1-J_2-J'$ lattice, where $J' = J_1$ describes the triangular lattice and $J' = 0$ describes decoupled honeycomb ($J_1-J_2$) and triangular ($J_2$) lattices.  The A and B sublattices of the honeycomb lattice and the C sublattice of central spins are labeled. (b) Unit cells: blue dotted lines show the small initial unit cell, while orange dashed lines show the larger final unit cell.  Both have trigonal symmetry - only the lattice vector changes. (c) These rotations convert the triangular lattice into the $J_1-J_2-J'$ lattice: the A and B clusters rotate in opposite directions, while the C clusters do not rotate. Inset shows original configuration. (d) The basic unit of the depleted fcc lattice: strong bonds are shown as red (solid) lines, weak bonds as blue (dashed) lines.  The central layer forms the emergent honeycomb lattice.}

\mot is a layered triangular lattice material built out of Mo$_3$O$_8$ clusters\cite{sheckelton12}.  Each cluster forms a molecular orbital with one Heisenberg spin-1/2 per three Mo.  The magnetic susceptibility follows a Curie-Weiss law within two different temperature regimes: a high temperature regime above 100K with Curie constant $C_H = .24$  emu K mol/Oe f.u. ($\mu_{H} = 1.39\mu_B$), corresponding to nearly the full $S=1/2$ moment and Weiss temperature, $\theta_H = -220K$; and a low temperature regime with Curie constant $C_L \approx 1/3 C_H$ and $\theta_L = -14K$.  This drastic moment reduction suggests that two-thirds of the spins vanish below 100K, which is consistent with the broad plateau in the entropy at $S \approx \frac{1}{3} R \log 2$ around 100K\cite{sheckelton12}.  Electron spin resonance measurements find the full $S=1/2$ moment (with $g = 1.9$) at low temperatures, confirming that this decrease is due to collective rather than single ion physics\cite{mcqueen12}. There are no sharp thermodynamic signatures, only a broad crossover in the susceptibility and a hump in the specific heat; Li NMR\cite{mcqueen12} and neutron\cite{sheckelton12} measurements have found no ordered moments, suggesting a gradual gapping out rather than a phase transition.  Sheckelton \emph{et al} proposed that the triangular lattice decouples into a valence bond solid (VBS) on the honeycomb lattice, with free central spins\cite{sheckelton12}.  However, if the lattice is really triangular, this decoupling is baffling - it should instead form a 120$^\circ$ ordered state\cite{huse88}.  To resolve this mystery, we propose that the triangular lattice physically distorts to favor this decoupling.

We suggest that the Mo$_3$O$_8$ clusters rotate as shown in Fig 1 (c), where clusters on the A and B honeycomb sublattices rotate in opposite directions, while the central clusters (C) do not rotate.  This rotation shortens the bond length between the honeycomb sites while lengthening that to the central spins.  In other words, the honeycomb nearest neighbor coupling, $J_1$  increases while the coupling to the orphan C spins, $J'$ weakens, favoring this decoupling.  This rotation can lead to a large change in bond length and thus $J$, due to the exponential dependence on the oxygen overlap. We parameterize this change to first order with $J_1 = (1+x)J_1^0$, $J' = (1-x)J_1^0$, where $x \in \{0,1\}$ smoothly interpolates between the triangular and honeycomb lattices. $J_2$ is unaffected.  The resulting $J_1-J_2-J'$ Hamiltonian is,
\begin{equation}
\label{heisenberg}
H = J_1\!\! \sum_{\langle ij\rangle_{A,B}}\!\!\vec{S}_i\cdot \vec{S}_j + J_2 \!\!\!\sum_{\langle\langle ij\rangle\rangle_{A,B}}\!\!\!\vec{S}_i\cdot \vec{S}_j + J'\!\!\!\!\!\!\!\!\!\! \sum_{\langle ij \rangle_{\{(A,B),C\}}}\!\!\!\!\!\!\!\!\!\vec{S}_i\cdot \vec{S}_j.
\end{equation}

We can gain a rough understanding by examining the variational energies of the triangular and honeycomb lattices.  Estimating $J_2/J_1$ from $\theta_L/\theta_H$ puts $J_2/J_1 \approx .06$[\onlinecite{sheckelton12}], well within the N\'eel region of the phase diagram [Fig. 2 (a)]. However, the low temperature regime is not magnetically ordered.  We shall show later that coupling to the orphan spins can drive the state towards the spin liquid, so we take the variational energy associated with the gapped spin liquid found for $J_2/J_1 \approx .06$, which is $-.5 J_1$ per honeycomb spin\cite{clark11}. At this point, we ignore the $J_2$ coupling of the orphan spins and treat them as free, making the energy per site $E_{\rm hex} = -.33 (1+x) J_1^0$. The triangular lattice energy is $-.537 J_1^0$ per site\cite{huse88}. The honeycomb and undistorted triangular energies cross at intermediate $x = .63$, although, as shown below we expect further corrections to favor the spin liquid. The lattice energy cost of the rotation will favor the triangular lattice, however we believe it is particularly small in this compound due to the cluster nature.  We have also neglected any intervening phases and a full numerical treatment should be done to get a more complete picture.

\fg{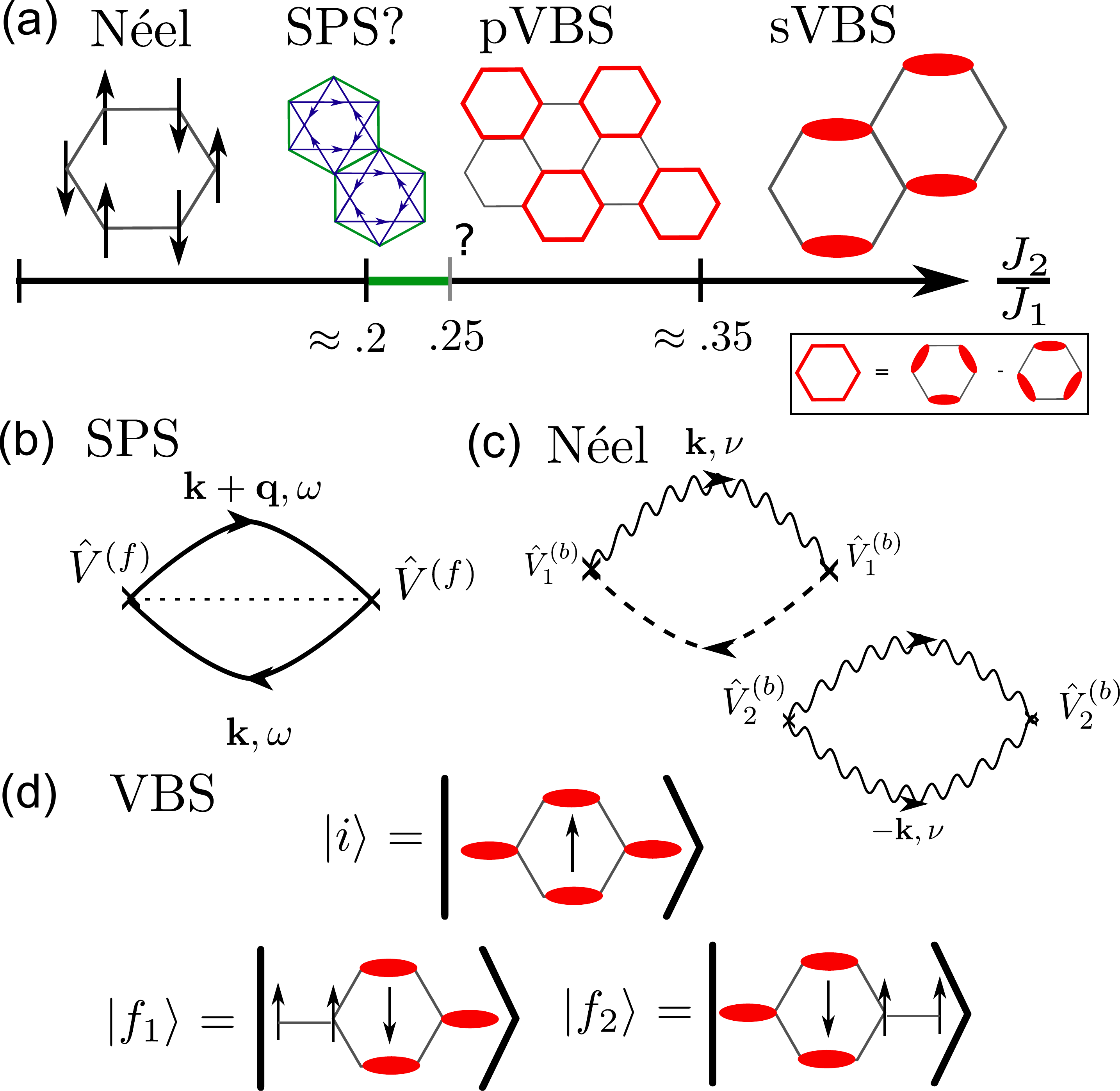}{variational}{(a) Rough phase diagram of the $J_1-J_2$ honeycomb lattice\cite{gong13}, with N\'eel, plaquette VBS (pVBS) and staggered VBS (sVBS) states, with a small controversial spin liquid region, thought to be the sublattice pairing state (SPS). (b) Diagram for the second order energy shift, $\Delta E^{(2)}_{\rm{SPS}}$ generated by a single central spin impurity in the SPS.  Solid lines are fermionic spinons, while the dashed line represents the central spin.  (c) Diagrams for the second order energy shift, $\Delta E^{(2)}_{\rm{AFM}}$ for the single central impurity in the N\'eel state. Squiggly lines represent Holstein-Primakoff bosons, $\alpha\dg_{{\bf k}}$, not magnons, and the dashed line are the Holstein-Primakoff bosons, $d\dg$ representing the central spin. (d) Initial and final spin configurations for calculating $\Delta E^{(2)}_{\rm{sVBS}}$, where the red ellipses represent singlet valence bonds.  Diagrams for $\Delta E^{(2)}_{\rm{pVBS}}$ are similar.}

How might these rotations be detected?  They triple the size of the unit cell [Fig. 1(b)], but leave the trigonal symmetry unchanged.  If the rotations form a static order, they should be seen with x-ray scattering.  So far this has not been found\cite{mcqueen12}, however they could instead be short range or even dynamic.  Short range order should be seen with further NMR or $\mu$SR measurements, but no matter the nature of the order, a soft phonon corresponding to these rotations should appear at the reciprocal lattice vectors of the honeycomb lattice.

In our variational picture, we left the central spins completely decoupled, both from the honeycomb lattice and from each other.  It turns out that these orphan spins favor the spin liquid over the competing N\'eel and VBS phases, as we shall now show by looking at a single central spin impurity in each of the four relevant phases.  The likely phase diagram of the $J_1-J_2$ honeycomb lattice is shown in Fig. 2(a).  Most studies\cite{clark11,albuquerque11,ganesh13,zhu13,gong13} agree that the N\'eel phase is stable below $J_2/J_1 \approx .2$ and that a staggered VBS (sVBS) is stable above $J_2/J_1 \approx .35$, but the middle of the phase diagram is more muddled.  There is a plaquette VBS below the sVBS, and there may be a narrow spin liquid region around $J_2/J_1 \approx .22-.25$[\onlinecite{gong13}], whose energy is consistent with the sublattice pairing state (SPS)\cite{gong13,lu11,wang10}.  This phase disappears quickly with either positive or negative $J_3$[\onlinecite{albuquerque11,gong13}], so the spin liquid region, if it exists, is clearly very narrow. All studies find a suprising second order phase transition between the N\'eel state and either the spin liquid\cite{gong13} or pVBS\cite{albuquerque11,ganesh13,zhu13}, suggesting deconfined criticality\cite{senthil04,wang10,moon12}.  

We begin with the SPS, which can be described with a fermionic spin representation with two spinons\cite{lu11,wang10}, $a_{i\sigma}$ and $b_{j\sigma}$ on the two sublattices.  The SPS is a mean-field state with a real nearest neighbor hopping amplitude, $t = \langle a\dg_{i\sigma} b_{j\sigma}\rangle$ (for $\langle ij\rangle$) and complex second neighbor pairing amplitudes with opposite phases on the two sublattices, $\Delta_A = \Delta \rm{e}^{i\theta} = \langle a\dg_{i\sigma} a\dg_{j-\sigma}\rangle$ and $\Delta_B = \Delta \rm{e}^{-i\theta} = \langle b\dg_{i\sigma} b\dg_{j-\sigma}\rangle$ (for $\langle\langle ij \rangle \rangle$)\cite{lu11}.  The mean-field Hamiltonian,
\bea
H & = & -t\! \sum_{\langle i j\rangle}\! \left[a\dg_{i\sigma} b_{j\sigma} + \rm{h.c.}\right]\! + \Delta\!\! \sum_{\langle\langle i j \rangle \rangle}\!\! \left[e^{i \theta} \mathrm{sgn}(\sigma)a\dg_{i\sigma}a\dg_{j-\sigma} \!+ \rm{h.c.}\right]\cr
& & + \Delta \sum_{\langle\langle i j \rangle \rangle} \left[e^{-i \theta} \mathrm{sgn}(\sigma)b\dg_{i\sigma}b\dg_{j-\sigma} + \rm{H.c.}\right]
\eea
can be diagonalized to give four bands,
\begin{equation}
\pm E_{{\bf k}\pm} = \pm \sqrt{t^2|\gamma_{\bf k}|^2 + \Delta_{\bf k}^2 \pm 2 t |\gamma_{\bf k} \Delta_{\bf k}| \sin \theta},
\end{equation}
where $\gamma_{\bf k} = 1 + \mathrm{e}^{i {\bf k} \cdot {\bf a}_2}+\mathrm{e}^{i {\bf k} \cdot {\bf a}_3}$; $\Delta_{\bf k} = \Delta (|\gamma_{\bf k}|^2 -3)$; and ${\bf a}$ are the real space lattice vectors: ${\bf a}_1 = {\bf x}$, ${\bf a}_2 = \frac{\bf x}{2} + \frac{\sqrt{3} \bf y}{2}$ and ${\bf a}_3 = {\bf a}_2-{\bf a}_1$ (where we have set the lattice constant to one).  For $\Delta = 0$, this dispersion is that of graphene, with Dirac cones at ${\bf K}$ and ${\bf K'}$.  These cones are gapped out by finite $\Delta$.  At special points, $\theta = 0, \pi/2$, there is a single gap minima and the spin liquid has a $U(1)$ symmetry.  For all other $\theta$'s, the SPS is a $Z_2$ spin liquid, with a line of minima around ${\bf K}$ and ${\bf K'}$ and a gap magnitude of $\Delta_{\rm g} = 6 \Delta$. Variational Monte Carlo results for $J_2/J_1 = .1$ find the best solution for $t = J_1$, $\Delta = .1 t$ and $\theta = 1$[\onlinecite{clark11}], which we use for the rest of the paper, although our results are relatively insensitive to these values, especially $\theta$.

As the SPS is a gapped spin liquid, it should be stable to the introduction of magnetic impurities up to $J' \sim \Delta_{\rm g}$.  In fact, the exchange coupling to the central spins is frustrated, increasing the stability.  The exchange coupling
\begin{equation}
\label{Hp}
J' \!\sum_{j = 1}^6 \vec{S}_j \cdot \vec{S}_7 = J'\! \sum_{\bk,\bq} \gamma_{A\bq} a\dg_\bk \vec{\sigma} a_{\bk+\bq}\cdot \vec{S}_7 + \gamma_{B\bq} b\dg_\bk \vec{\sigma}b_{\bk+\bq} \cdot \vec{S}_7,
\end{equation}
where $\gamma_{A\bq} = {\rm e}^{i \ba_1 \cdot \bq} \gamma_\bq$, $\gamma_{B\bq} = {\rm e}^{i \ba_2 \cdot \bq} \gamma^*_\bq$ is reminiscent of a Kondo coupling to spinons instead of electrons.

We fix the central spin and calculate the energy shift to second order with the diagram in Fig. 2(b):
\bea
\Delta E^{(2)}_{\rm SPS} & = & |J'|^2 \sum_{\bk,\bq} T\sum_{i\omega_n} |\hat V_{\bk,\bk+q}^{\eta_1 \eta_2}|^2 \mathcal{G}_{\eta_1}(\bk+\bq,i\omega_n) \mathcal{G}_{\eta_2}(\bk,i\omega_n) \cr
& = & |J'|^2 \sum{\bk,\bq} |\hat V_{\bk,\bk+q}^{\eta_1 \eta_2}|^2\frac{f(E_{\bk+\bq,\eta_1}) - f(E_{\bk,\eta_2})}{E_{\bk+\bq,\eta_1} - E_{\bk,\eta_2}}\cr
& = & -2 |J'|^2 \sum_{\bk,\bq}\sum_{\eta_1 \in {\rm c}, \eta_2 \in {\rm v}} \frac{|\hat V_{\bk,\bk+q}^{\eta_1 \eta_2}|^2}{E_{\bk+\bq,\eta_1} + E_{\bk,\eta_2}},
\eea
where we take $T \rightarrow 0$, fixing $\eta_1$ and $\eta_2$ in the conduction and valence bands, respectively.
The matrix element is:
\bea
\hat V_{\bk,\bk+q}^{\eta_1 \eta_2} & = & \gamma_{A\bq}\left[U\dg_\bk\left(\begin{array}{cc}\tau_+ & 0\\ 0 & 0\end{array}\right) U_{\bk+\bq} \right]^{\eta_1 \eta_2}\cr
& & + \gamma_{B\bq}\left[U\dg_\bk \left(\begin{array}{cc}0 & 0\\ 0 & \tau_+ \end{array}\right) U_{\bk+\bq} \right]^{\eta_1 \eta_2},
\eea
where $U_{\bk}$ is the eigenvector matrix diagonalizing the Hamiltonian: $U_{\bk}\dg H_{\bk} U_{\bk} = \rm{diag}(E_{\bk\eta})$.  Upon numerical integration, we find:
\begin{equation}
\Delta E^{(2)}_{\rm SPS} =-3.4\frac{|J'|^2}{J_1}.
\end{equation}

The energy shift in the VBS phases can be calculated directly through second order perturbation theory,
\begin{align}
\Delta E^{(2)}_{\rm VBS} & = -\sum_n
\frac{|\langle i| H' |f_n\rangle|^2}{\Delta_{\rm s}}
\end{align}
by applying $H' = J' \sum_{j = 1}^6 \vec{S}_j \cdot \vec{S}_7$ to representative hexagons, as shown in Fig. 2(d) for the sVBS phase, where two intermediate states $|f_1\rangle$ and $|f_2\rangle$ are generated from $H'|i\rangle$.  $\Delta_s =  \frac{3}{4}|J_1|$ is the singlet gap. The pVBS state proceeds similarly, but with more book-keeping. Notably, $H'$ kills the plaquette singlet states shown in the inset of Fig 2(a) and the pVBS state gains the least energy from the orphan spins.  The two energy shifts are:
\begin{align}
\Delta E^{(2)}_{\rm sVBS} = -\frac{4|J'|^2}{3J_1} = -1.3 \frac{|J'|^2}{J_1}\cr
\Delta E^{(2)}_{\rm pVBS} = -\frac{2|J'|^2}{3J_1} = -.67 \frac{|J'|^2}{J_1}.
\end{align}

Finally we calculate the energy shift in the N\'eel phase using spin wave theory.  We introduce three Holstein-Primakoff bosons\cite{hp} for the three sublattices: $a$, $b$ and $d$.  As the spins on the A and B sublattices are antiparallel, we rotate the B spins, while fixing C parallel to A:
\bea
S^z_{A} & = & S-a\dg a, S^+_A = \sqrt{2S}a\dg, S^-_A = \sqrt{2S}a\cr
S^z_{B} & = & -S+b\dg b, S^+_B = \sqrt{2S}b, S^-_B = \sqrt{2S}b\dg\cr
S^z_{C} & = & S-d\dg d, S^+_C = \sqrt{2S}d\dg, S^-_C = \sqrt{2S}d.
\eea
To $O(S)$, the honeycomb spin-wave Hamiltonian is
\bea
H_0 & = & -3 J_1 \mathcal{N}_s S^2 + J_1 S\sum_{\bk} 3 a_\bk\dg a_\bk + 3 b\dg_\bk b_\bk \cr
& & +\left[\gamma_\bk^* a\dg_\bk b\dg_{-\bk} + \gamma_\bk b_\bk a_{-\bk} + \rm{H.c.}\right]\cr
& = & -3 J_1 \mathcal{N}_s S^2 + \sum_{\bk \eta} \omega_{\bk \eta}\left(\alpha\dg_{\bk\eta}\alpha_{\bk\eta} + \frac{1}{2}\right),
\eea
where $\mathcal{N}_s$ is the number of sites, $\gamma_\bk$ is defined above and the dispersion $\omega_{\bk \eta} = \omega_{\bk} = J_1 S\sqrt{9-|\gamma_\bk|^2}$ is identical for the two bands of magnons.  $\omega_{\bk}$ has minima at $0$, ${\bf K}$ and ${\bf K'}$, as expected. The Hamiltonian is diagonalized by
\bea
\label{diag}
\alpha\dg_{1\bk} & = & u_\bk b\dg_\bk - v_\bk^* a_{-\bk}\cr
\alpha\dg_{2\bk} & = & u_\bk^* a\dg_\bk + v_\bk b_{-\bk}.
\eea
The impurity interaction is given by
\bea
H' & = & J'S \sum_\bk \left[\gamma_{A\bk} a\dg_\bk d + \gamma_{B\bk} b\dg_{\bk} d\dg +\rm{H.c.}\right]\cr 
& &+ J'S\sum_{\bk,\bq} \gamma_{B\bq} b\dg_\bk b_{\bk+\bq} -\gamma_{A\bq} a\dg_\bk a_{\bk+\bq}.
\eea
We can calculate the second order energy shift directly with second order perturbation theory, or equivalently with the diagrams in Fig 2(c). The particular structure of the impurity interaction means that the original $\alpha$ bosons and not the honeycomb magnons, $\alpha + \alpha\dg$ are the relevant particles.  There are two intermediate states: spin flip scattering off the impurity state, $|f_{1\bk}\rangle = \alpha\dg_{\eta \bk} d\dg |0\rangle$ and potential scattering off the impurity spin that creates two bosons, $|f_{2\bk,\bk'}\rangle = \alpha\dg_{\eta \bk}\alpha\dg_{\eta' \bk'}|0\rangle$.  The matrix elements are:
\bea
\!\!\!\!\!\!\!\!V_{1\bk} & = & \langle f_{1\bk}| H' |0\rangle = J' S\left(u_\bk \gamma_{B\bk}-v_\bk^*\gamma_{A\bk}\right)\cr
\!\!\!\!\!\!\!\!V_{2\bk,\bk'} & = & \langle f_{2\bk,\bk'}| H' |0\rangle = J'S \!\left(u_{\bk'} v_\bk \gamma_{A \bq}- u_\bk v_{\bk'}^* \gamma_{B \bq}\right)\!,
\eea
where we have used Eqn (\ref{diag}), and defined $\bq \equiv \bk'-\bk$, leading to the overall energy shift
\bea
\Delta E_{\rm AFM}^{(2)} & = & -\sum_\bk\frac{|V_{1\bk}|^2}{\omega_\bk} - \sum_{\bk,\bk'} \frac{|V_{2\bk,\bk'}|^2}{\omega_{\bk} + \omega_{\bk'}}\cr
 & = & -1.2 \frac{|J'|^2}{J_1}.
\eea

So we have found that the SPS energy shift is more than twice as large as either the N\'eel or VBS energy shifts, indicating that the central spins are stabilizing the spin liquid state over the other states - a simple interpretation is that in the spin liquid state, the hexagon surrounding the orphan spin can gain more energy by being polarized as compared to the N\'eel or VBS states. 

How can these three states be experimentally differentiated? Symmetry-wise, the spin liquid breaks no symmetries while the N\'eel state breaks spin rotational symmetry, the sVBS breaks trigonal symmetry (as the dimers select a direction) and the pVBS breaks translation symmetry to enlarge the unit-cell three-fold. The magnetic excitations, as measured by inelastic neutron scattering (INS) also differentiate between the VBS phases and the SPS.  In a VBS, the spinons are confined, which gives a sharp, nearly dispersionless singlet-triplet excitation at $\Delta_{\rm s} = 3/4 |J_1| \approx 165$K.  The SPS spinons are also gapped, with a similar magnitude ($\Delta_{\rm g} = .6 |J_1|$) but deconfined, so the INS signal turns on gradually above the gap and the excitation has a fairly large bandwidth of $6|J_1|$.  We have done an RPA calculation to extract the power law behavior at gap minima, $\bq = 0$ and $\bq = {\bf K}-{\bf K'}$, where we find $\chi''(\omega = \Delta_{\rm g} + \delta \omega)$ turns on as $\delta \omega^2$ and $\delta \omega$, respectively (see supplementary material).  The naive expectation is that a convolution of spinons yields a step function at the threshold, but matrix elements cause further suppression and we expect a highly smeared spectral function.

The idea of emergent lattices provides a powerful new way to stabilize spin liquids, and there is an intriguing possible 3D application in Ba$_2$YMoO$_6$, which also shows two distinct paramagnetic regimes\cite{aharen10,devries10}.  Here, the Mo sit on an fcc lattice, which is expected to order magnetically.  However, these Mo$^{5+}$ ions have one 4d electron in the three-fold degenerate t$_{2g}$ orbital, forming a $J_{\rm eff} = 3/2$ \emph{quartet}.  The $J_{\rm eff}$ quartet is unusual in that it has no intrinsic magnetic moment\cite{kotani49}, although hybridization with the surrounding oxygens can restore it\cite{chen10}, explaining the experimentally observed moment.  There is no sign of the expected Jahn-Teller distortion down to 2K, and the full $R \log 4$ entropy is recovered by 150K\cite{devries10}, implying that the moments retain their full $SU(4)$ symmetry.  While the Hamiltonian will not be $SU(4)$ symmetric without engineering, the absence of ordering suggests that the $SU(4)$ quantum fluctuations might increase the chance of finding a spin liquid. The quartet nature means that a singlet involves four sites: a singlet valence plaquette\cite{li98,pankov07,xu08}, and opens up a rich possible phase diagram.  

So Ba$_2$YMoO$_6$ can be described as $J_{\rm eff}=3/2$ quartets on an fcc lattice, which, like the triangular lattice in \motp, should order.  And yet there is no sign of a phase transition in the thermodynamic measurements, or of magnetic order in $\mu$SR\cite{devries10}, neutron\cite{carlo11} or Y NMR\cite{aharen10} measurements.  Instead, the susceptibility shows a high temperature Curie-Weiss regime, with $C_H \approx .25$ emu/mol K and $\theta_H = -160$K that crosses over around 50K to a low temperature Curie-Weiss regime with $C_L \approx .15 C_H$ and $\theta_L = -2.3$K\cite{devries10}, suggesting that around 85\% of the Mo spins vanish below 50K.

The fcc lattice can be thought of as an ABC stacking of triangular lattices; this structure naturally suggests a construction of emergent lattices by decoupling one of the layers into a honeycomb lattice with weakly coupled central spins - see Fig 1 (d) for the basic unit.  If we decouple every other layer, creating an $AB'CA'BC'$ stacking, where $N'$ indicates a decoupled layer, we create a \emph{one-sixth depleted fcc lattice}.  If the strongly coupled spins form a valence bond solid or spin liquid, 17\% orphan spins will remain at low temperatures, close to the number seen in Ba$_2$YMoO$_6$.  However, as each Y atom will have exactly one orphan spin as a neighbor, this decoupling cannot explain the development of two Y NMR sites below 50K\cite{aharen10}.  Another possibility is to decouple every third layer, $ABC'ABC'$, forming a \emph{one-ninth depleted fcc lattice} that leaves behind 11\% orphan spins.  This lattice creates two Y NMR sites: those with one neighbor orphan spin and those with none, in a two to one ratio.  Both possibilities expand the original fcc unit cell to a six- or nine-fold larger hexagonal unit cell, which could be detected with x-ray scattering. 
Ba$_2$YMoO$_6$ seems to have 15\% orphan spins, intermediate between these two options.  So another possibility is that the depleted layers are randomly distributed and the average spacing is between one and two layers. This arrangement would still lead to two Y NMR sites, but would not show up in x-ray measurements.  However, as with \motp, we expect some kind of lattice distortion to favor this decoupling and there should be a corresponding soft phonon around 50K. While a spin liquid would be the most exciting possibility for these depleted fcc lattices, inelastic neutron measurements detect a sharp excitation at 28meV that looks like a singlet-triplet gap\cite{carlo11}, suggesting a VBS or plaquette solid instead.

In this paper, we have suggested the formation of emergent lattices weakly coupled to the remaining spins as an explanation for the two distinct paramagnetic regimes in both \mot and Ba$_2$YMoO$_6$.  However, this idea is much more general and provides a novel mechanism to stabilize quantum spin liquids in both two and three dimensions. Future numerical and theoretical work should examine this idea in more detail on specific lattices and check the full phase diagram.  Experimentally, this idea may be useful in attempts to engineer spin liquid materials via creating artificial emergent lattices.

{\bf Acknowledgements:} The authors thank Bryan Clark, Hong-Chen Jiang, Tyrel McQueen and John Sheckelton for useful discussion. RF is supported by the
Simons Foundation and PAL is supported by NSF DMR 1104498.

\section*{Supplementary material: RPA calculation of the SPS structure factor}

Here, we calculate the power law dependence of the structure factor, $S({\bf q},\omega) = \frac{1}{\pi} (n(\omega)+1) \chi''(\bq,\omega)$ in the SPS just above the gap, $\Delta_{\rm g}$ near both $\bq = 0$ and the antiferromagnetic wave vector, $\bq = {\bf Q} \equiv {\bf K} - {\bf K'}$.  For simplicity, we approximate the $\theta = 1$ dispersion with the $\theta = \pi/2$ limit, which has a single minima at ${\bf K}$ and ${\bf K'}$ rather than a line of minima around each of those points.  The dispersion,
\begin{equation}
E_{\bk \eta} = \eta \sqrt{t^2|\gamma_\bk|^2+\Delta_\bk^2}
\end{equation}
is four-fold degenerate after including spin.  We calculate
\begin{equation}
\chi''(\bq,\omega) \propto \int \frac{d \nu}{2\pi} \sum_\bk \sum_{\alpha,\beta \in (A,B)} \sum_n \mathcal{G}_m^{\alpha \beta}(\bk+ \bq,\nu + \omega) \bar{\mathcal{G}}_m^{\beta \alpha}(\bk,\nu),
\end{equation}
where $\mathcal{G}_m$ is shorthand for the normal, $G$ and anomalous, $F$ Green's functions.  Generically the Green's functions for A and B spinons can be written as,
\begin{equation}
\mathcal{G}_m^{\alpha \beta}(\bk, i \omega_n) = \sum_\eta \frac{C_{\eta m}^{\alpha \beta}(\bk)}{i \omega_n - E_{\bk \eta}} +  \frac{D_{\eta m}^{\alpha \beta}(\bk)}{i \omega_n + E_{\bk \eta}}.
\end{equation}
For $T = 0$ and $\omega > 0$, $\chi''(\bq,\omega)$ becomes:
\bea
\chi''(\bq,\omega) & = &  -\frac{1}{\pi}\!\! \sum_{\bk,\eta m,\alpha \beta}\!\!\!\! C_{\eta m}^{\alpha \beta}(\bk+\bq)D_{\eta m}^{\beta \alpha}(\bk)  \delta\left[ \omega - E_{\bk+\bq} - E_\bk\right]\cr
 & = & -\frac{1}{\pi} \!\sum_{\bk}\!\!\left(\!\!1-\frac{t^2\!\!\left[\gamma_\bk^* \gamma_{\bk+\bq} + \gamma_{\bk+\bq}^* \gamma_\bk\right]\! + 2 \Delta_{\bk+\bq}\Delta_\bk}{2E_{\bk+\bq} E_\bk}\!\right)\cr
& & \times \delta\left[\omega - E_{\bk+\bq}-E_\bk\right].
\eea
We now make several more simplifying assumptions: that $\omega = \Delta_{\rm g} + \delta \omega$ is close to the gap edge; that $\bq = \bq_0 + \delta \bq$ is very close to $\bq_0 \in ({\bf 0}, {\bf Q})$; that $\bk = {\bf K}_0 +\delta \bk $ is close to one of the minima at ${\bf K}_0 \in({\bf K},{\bf K})$; and finally that the dispersion is spherically symmetric around ${\bf K}_0$ instead of trigonal.  These mean that $E_{\bk +\bq} + E_\bk \approx \Delta_{\rm g} + \frac{\delta \bk^2}{2m}+ \frac{(\delta \bq + \delta \bk)^2}{2m}$.  We also define $\delta \bk$ such that $\delta \bq = (\delta q, 0)$, which allows us to write out the matrix elements, $A(\bk,\bq) = \sum_{\eta m, \alpha \beta}  C_{\eta m}^{\alpha \beta}(\bk+\bq)D_{\eta m}^{\beta \alpha}(\bk)$ for $\bq_0 = {\bf 0}, {\bf Q}$:
\begin{align}
\!\!\!\!\!\!\!\!A_{\bf 0}(\delta \bk, \delta q) = & \frac{25}{6}\delta q^2 + O(\delta q^3,\delta {\bf k} \delta q^2) \cr
\!\!\!\!\!\!\!\!A_{\bf Q}(\delta \bk, \delta q) = & \frac{25}{6}\delta q^2 + \frac{50}{3}(\delta q \delta k_x +\delta k_x^2) + O(\delta[k,q]^3)
\end{align}
$A_{\bf 0}$ actually vanishes for $\delta q = 0$, independent of $\delta \bk$.  Performing the k-integral analytically, we find:
\begin{align}
\chi''(\delta\bq,\Delta_{\rm g }\! +\! \delta\omega)\! = &  \frac{25}{12} \delta q^2 + O(\delta \omega^2\!\!,\delta q^3)\cr
\!\!\!\!\!\!\chi''({\bf Q}\! +\! \delta\bq,\Delta_{\rm g }\! +\! \delta\omega)\! = & \frac{125}{48} \delta q^2\! +\! \frac{25 m \delta \omega}{24}\! +\! O(\delta \omega^2\!\!,\delta q^3\!).
\end{align}
where $m = 20/47$ and everything is in units of $t = J_1$.  

The main takeaway here is that the neutron form factor for the SPS is very smeared out in momentum space, turning on above the gap only as $\delta \omega^2$ for $\bq = 0$ and as $\delta \omega$ for $\bq = {\bf Q}$, while the VBS singlet-triplet gap is much sharper.  For small $\delta \omega$ and $\delta \bq$, the approximation of a single minima fails and we expect one-dimensional behavior due to the line of minima, however the qualitative picture of a smeared out spectrum will hold.


\end{document}